# Reviewing Literature on Time Pressure in Software Engineering and Related Professions

Computer Assisted Interdisciplinary Literature Review

Miikka Kuutila, Mika V. Mäntylä, Maëlick Claes
M3S / ITEE
University of Oulu, Oulu, Finland
{miikka.kuutila, mika.mantyla, maelick.claes}@oulu.fi

Marko Elovainio
Department of Psychology
University of Helsinki, Finland
Marko.Elovainio@helsinki.fi

*Abstract* — **During the past few years, psychological diseases related to unhealthy work environments, such as burnouts, have drawn more and more public attention. One of the known causes of these affective problems is time pressure. In order to form a theoretical background for time pressure detection in software repositories, this paper combines interdisciplinary knowledge by analyzing 1270 papers found on Scopus database and containing terms related to time pressure. By clustering those papers based on their abstract, we show that time pressure has been widely studied across different fields, but relatively little in software engineering. From a literature review of the most relevant papers, we infer a list of testable hypotheses that we want to verify in future studies in order to assess the impact of time pressures on software developers' mental health.**

***Keywords; Software engineering; time pressure; speed-accuracy tradeoff; Job demands-resources model, deadline pressure, topic analysis***

## I. Introduction

Time pressure has been shown to associate with arousal, negative emotions and perceived stress, these in turn have some strong physiological correlates, such as an increased heart rate and hormonal responses. Stress also intensifies all other emotions [1], e.g. frustration can change to anger. Several theories on the outcomes of time pressure on performance and motivation exist. Among them are Yerkes Dodson's law [2], Job demands-resources model [3] and Speed-accuracy tradeoff [4].

Yerkes Dodson's law [2] dictates a U-shaped relation of time pressure and performance. Performance increases with arousal, but when arousal lasts too long, or the level of arousal is too high, performance starts to decrease. Moreover, performance decreases without arousal. In software context, this means deadlines can increase the performance of developers, but when the requisites for deadlines become too high the performance decreases.

Job demands-resources model [3] generalizes every job as having demands, often defined as time pressure, and resources. It examines the relationship between job demands and resources and their effects on strain and motivation. The model assumes that job strain is caused by imbalance between resources and demands, such as lack of time or too much demand before the next deadline.

Speed-accuracy tradeoff [4] is a phenomenon observed among humans and animals alike. It has been observed that decision speed covaries with decision accuracy. Simply put, quickly made decisions contain more errors, which in software development context might prove costly.

Software development offers data for researchers, which may be lacking elsewhere: analyzable data of work performance exists in software repositories in the form of commit and chat logs. Observing and predicting time pressure could be used to maintain optimal time pressure and avoid burnouts, thus improving the job wellbeing of individual developers and boosting their performance per Yerkes Dodson's law.

This study aims to explore the literature related to time pressure by performing a computer assisted interdisciplinary literature review. Our approach is motivated on our observation that time pressure has been studied in multiple disciplines and hundreds of papers about the topic have been published. Our goal is to offer motivation and theoretical background for a time pressure detection framework specific to the field of software development. Additionally, some literature from other fields is reviewed to see whether lessons from elsewhere could be useful in a software development context.

The remainder of the paper is structured as follows. First Section II presents the general methodology we used to extract and analyze 1270 papers found on Scopus and related to time pressure. In Section III we present the result of the clustering of those papers' abstracts. In Section IV we review the most relevant papers related to software engineering, but also from other fields from which we can potentially gain knowledge that can be transposed to software engineering. In Section V, we discuss potential hypotheses found in the literature and how they could be studied in software repositories. Finally, the paper concludes by presenting our planned future work.

## II. Methodology

Scopus database was used to gather information on papers related to time pressure. Scopus provided all information used in analysis, including titles, abstracts and citations. The data from Scopus was collected on 9th November 2016 with the search string *“(TITLE ("time pressure") OR KEY ("time pressure") OR TITLE ("schedule pressure") OR KEY ("schedule pressure") OR TITLE ("time budget pressure") OR KEY ("time budget pressure") OR TITLE ("deadline pressure") OR KEY ("deadline pressure") OR TITLE ("pressure of time") OR KEY ("pressure of time") OR TITLE ("pressure of schedule") OR KEY ("pressure of schedule") OR TITLE ("pressure of time budget") OR KEY ("pressure of time budget") OR TITLE ("pressure of deadline") OR KEY ("pressure of deadline") OR TITLE ("speed-accuracy tradeoff") OR KEY ("speed-accuracy tradeoff") AND NOT TITLE-ABS-KEY ("long rise-time") AND NOT TITLE-ABS-KEY (intracranial) AND NOT TITLE-ABS-KEY (drill) AND NOT TITLE-ABS-KEY (space-time))*”. The search string contains several search terms and different ways of spelling them, all related to time pressure. Additional search terms were found by familiarizing with the search results and improving the search string incrementally. The search from Scopus yielded a total of 1378 papers. This initial dataset was further filtered both by hand and programmatically, bringing the amount of papers down to 1270 papers. This was done by searching for entries containing suspect strings (e.g. “time, pressure” and “real-time pressure”) and reading them through. This resulted in finding papers from the data that deal with thermodynamics, which were clearly out of scope and thus were excluded.

The filtered data was further analyzed with Excel and R. Clustering of the data was done using R package *topicmodels*, using latent Dirichlet allocation (LDA) and Gibbs sampling. This means that each paper is only assigned to one cluster or topic, the approach has been introduced in a widely cited paper by Griffiths and Steyvers[5]. Optimal number of topics was selected using a log-likelihood measure, which provided 79 topics.

## III. Results of Clustering

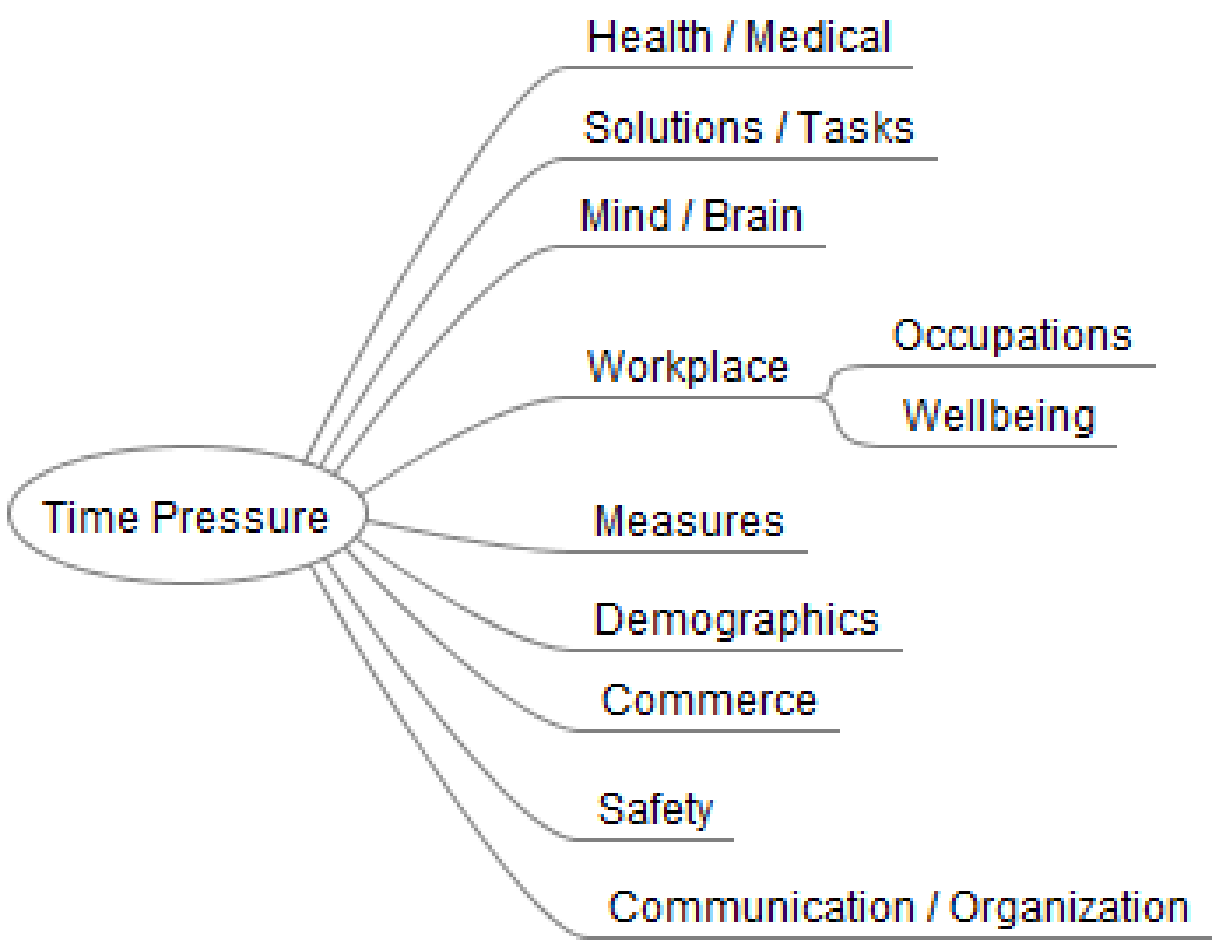

Figure 1. High level of qualitative coding

The acquired 79 topics were qualitatively encoded using FreeMind tool, further clustering the topics and making them easier to perceive. Qualitative coding was assisted by examining the papers each topic contains. The highest level of the qualitative coding, depicted in Figure 1, contains nine different nodes: Safety, Workplace, Solutions / Tasks, Commerce, Communication / Organization, Mind / Brain, Measures, Demographics and lastly Health / Medical. One topic can be under several nodes.

Node Safety has five different topics and are related to accidents, emergencies and states and substances influencing behavior e.g. fatigue and alcohol. Workplace is divided into two nodes, Occupations containing 12 different topics and Wellbeing containing 5 topics. Wellbeing focuses on the effects of work with probable words such as job satisfaction, burnout, and emotional wellbeing as well as working patterns: maintenance, day, pattern and shift. Node Solutions/Tasks has 10 different topics related to studied tasks such as visual searching and solutions such as adaptive keyboards. Six different topics under node Commerce are related to commercials, shopping, products and finances while nine topics under node Communication / Organization are related to words such as groups, negotiations and collaboration. Node Mind / Brain is divided into several nodes under it: brain functions, mental states, decision-making and behaviour. Measures contain two topics with words such as pupil, diameter and stimulus. Node demographics contains 5 topics with words that are demographic variables, such gender, elderly, adults and children. Lastly, node Health / Medical contains five different topics with words related to the medical field such as controls, patients, symptoms, food, consumption, services, medical and care.

## IV. Overview of Topics Related to workplace

The clustering yielded several topics related to occupations which can be viewed in Figure 2. In addition to reviewing papers under topic software engineering, we review relevant papers on topics project work and auditing that we consider related to software engineering, as those areas can have findings relevant to software engineering. Project work is part of software projects, while auditing may be considered to require similar attention to detail as programming. We also review papers that were clustered to topic general job satisfaction, as they are certainly relevant to software engineering.

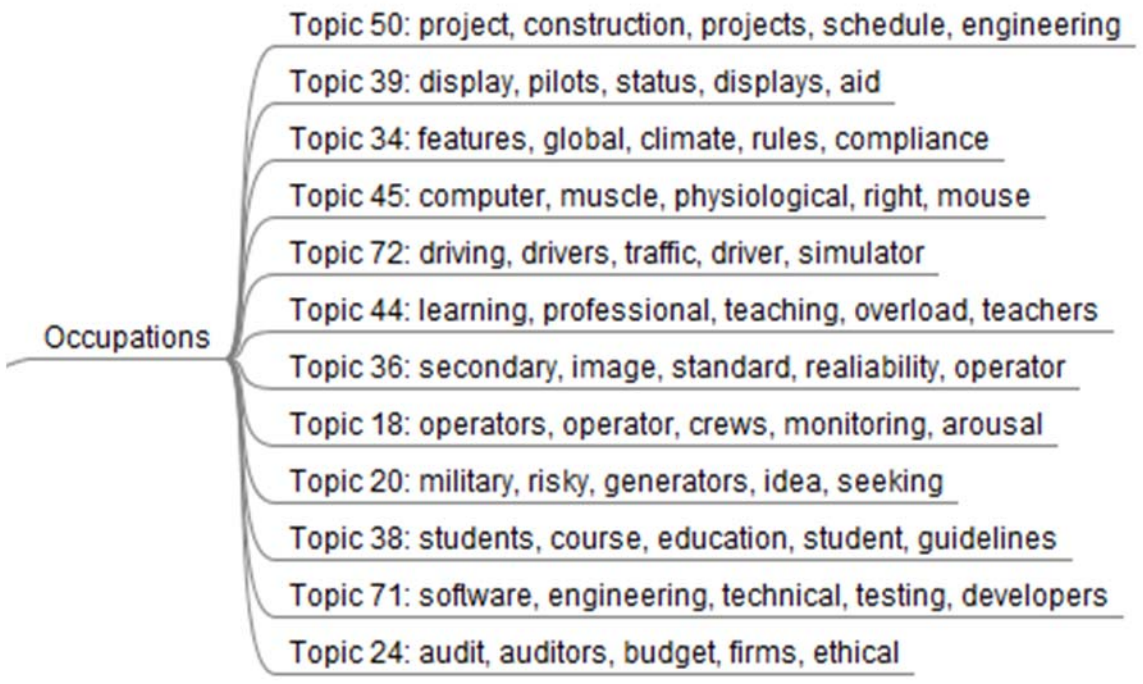

Figure 2. Topics belonging to the node “Occupations”.

### A. *Software engineering*

Clustering the papers provided a topic with the five most probable words “software”, “engineering”, “technical”, “testing” and “developers”. Many of the 21 papers included in this topic take a closer look on several software development processes in the organizational level, such as startups or agile development. Comparing the word-cloud of Figure 3 with word-clouds in Figure 4-6, words time and pressure are noticeably smaller in Figure 3. This means that these words are used less in the titles and abstracts of papers, implying that time pressure is less studied in software engineering.

After familiarizing with the papers of this topic, most of the papers are not directly focused on time pressure on software engineering. Instead it is mentioned as an effect contributing to the examined phenomenon (e.g. technical debt, code quality, pivoting). Interestingly, six out of twenty-one papers in this cluster are concerned with testing and code quality, which could indicate that time pressure has more impact in these steps of the development process.

Nan and Harter [6] have done a literature review on resource constraints on software development and a case study examining the relationship between pressure and performance with both budget and schedule pressure. The data used in the case study comes from large company developing information systems for both government and private companies. The findings support Yerkes Dodson’s law, both pressures caused by schedule and budget pressure were found out to have “U-shaped” relationship with development effort, i.e. having too much or too little pressure lead to less development effort. The paper stands out of this cluster for the combination of an empirical case study and its focus on resource scarcity and time pressure.

Harris, Collins and Hevner [7] focused on the role of deadlines in agile development. The whole underlying assumption in agile methods is to deliver usable software fast, while small teams and weekly project meetings make slack visible. The authors noted that these weekly meetings act as deadlines, and increase both motivation and strain. Such findings are indeed supported by psychological studies showing how deadlines

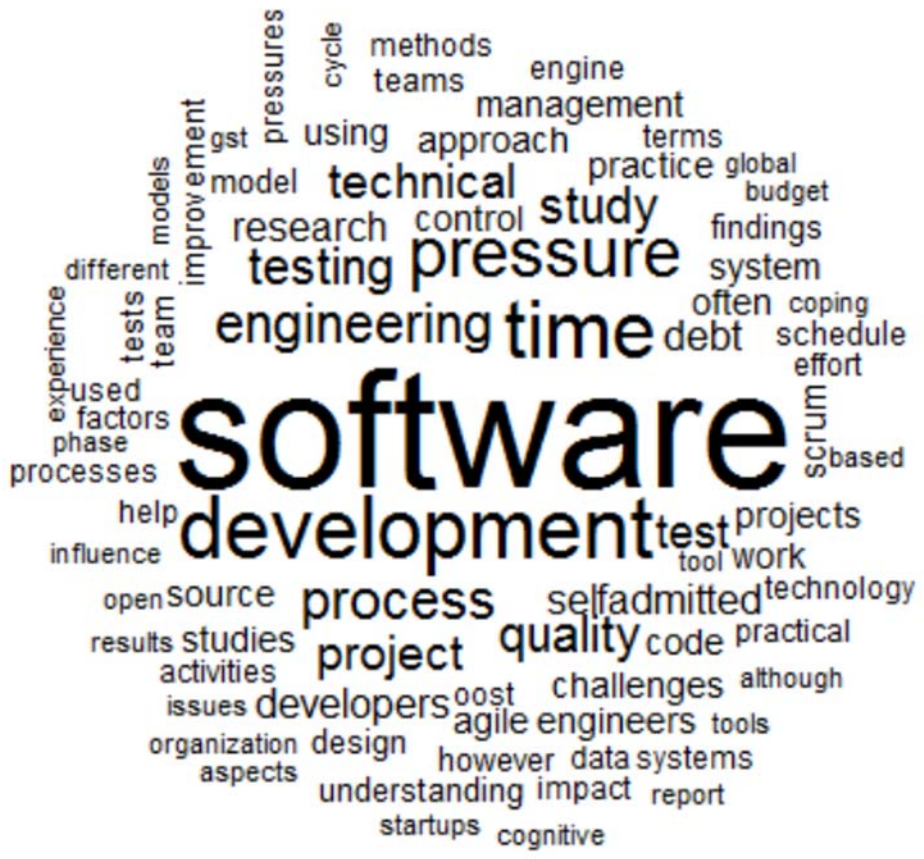

Figure 3. Wordcloud of Topic “Software engineering” made from the words of title and abstract

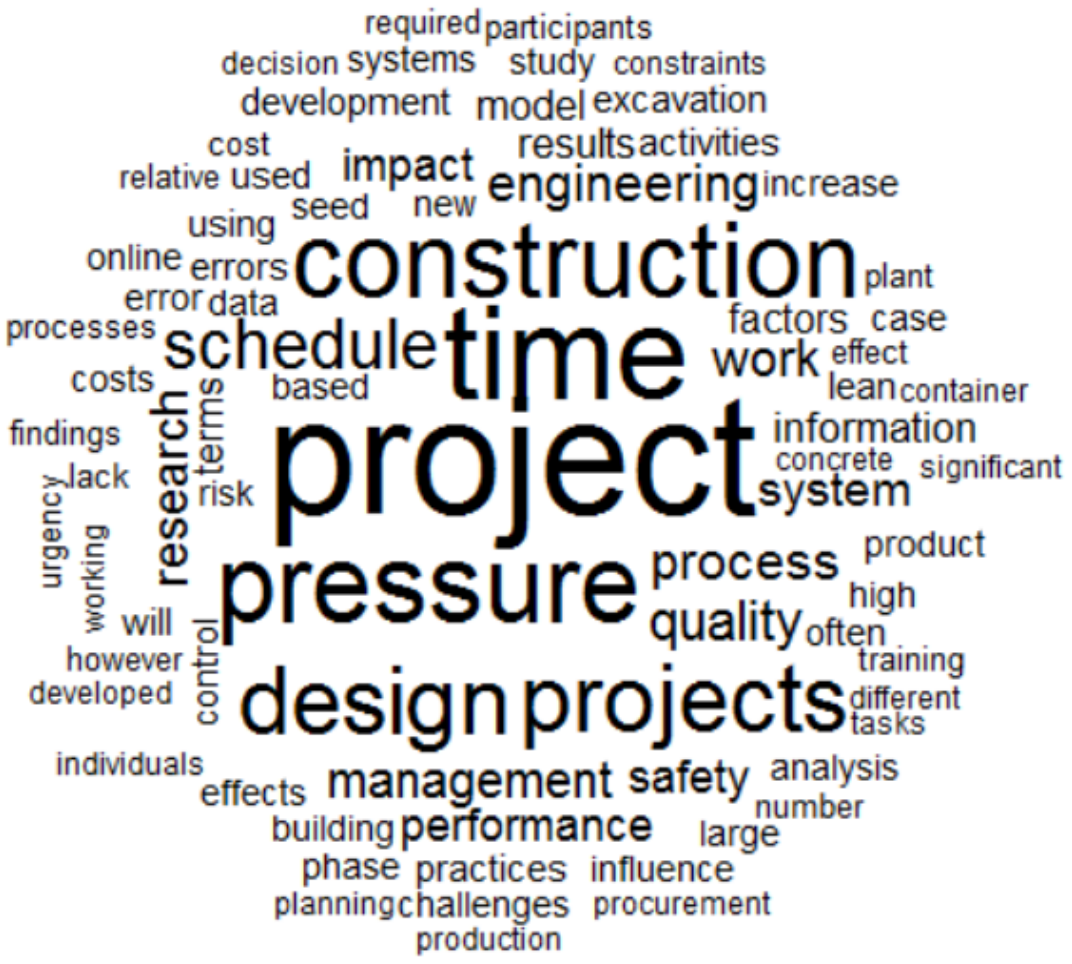

Figure 4. Wordcloud of topic “Projects”, made from titles and abstracts.

help to avoid procrastination and improve productivity and quality, e.g. [8] .

Hazzan and Dubinsky [9] study time management in software context and they perceive time pressure as a time-related problem along with concepts such as “bottlenecks”, “project-planning and schedule”, “Time estimation” and “late delivery” related to time management. After a literature review, they also noted that attention to time related problems has historically increased in software engineering. Authors also suggested that difficulties in time management in software context could be caused by its abstract nature and the limitation of human senses to gauge it. Additionally, authors note that tight development processes improve time-management as they systemically support the tracking of time.

### B. *Project work*

The five most probable words of this topic are “project”, “construction”, “projects”, “schedule” and “engineering”. Many of the 30 papers clustered are related to construction projects, but papers from all kinds of projects are included, meaning there were a few for Information Systems as well.

Case study by Van Berkel, Ferguson and Groenewegen [10] found out that pressure by political stakeholders for project completion can negatively impact public projects, as well as public scrutiny. Additionally, it was noticed that communication and coordination between temporal and permanent organizations was hampered by the pressure for completion.

Hussein [11] used gaming simulation to research project management. Even though the main goal of the research presented was to demonstrate the usefulness of gaming simulation as a research method, some useful observations were made regarding time pressure. On the early phases of a simulated project, time pressure or “urgency” lead to response patterns: “tendency to overly focus on the technical solution”, “tendency to make unverified assumptions” and lastly “significant rise to personal emotions, such as fear, diffidence, competitiveness and eagerness”.

Cataldo [12] analyzed data from over 200 software projects by a globally distributed organization to see common causes for errors and implications for the use of collaborative tools. The main outcome variable was the number of defects reported during integration and testing, while time pressure was computed as the standard deviation of the number of completed tasks in each month. He found out that most significant source of errors were affected by time pressure and uneven distribution of workers across locations. Other sources, which paled in comparison, were identified as the technical experience of project members and technical dependencies of the project.

Interestingly, the topic also contained one paper related to information systems dating back to 1994 [13], which highlights that researchers interested in human factors and information systems were already interested in time pressure twenty years ago. Hwang [13] suggested a model that defines how time pressure should impact decision performance, decision strategy, task difficulty, goal commitment and information systems based on studies in other fields.

### C. *Auditors and Auditing*

Similarities in the work of auditors and software developers can be found as both jobs require attention to the detail and understanding of rules. Hence it was decided to overview this topic made with clustering. This topic's five most probable words are "audit", "auditors", "budget", "firms" and "ethical" and it consists of 26 papers in total.

Job performance in relation to time pressure has been extensively researched in auditing. Broberg, Tagesson, Argento, Gyllengahm and Mårtensson [14] examined with surveys whether "time budget pressure" influenced audit quality in Sweden. Negative relationship between audit quality and "time budget pressure" was indeed noted, additionally having more clients also impacted audit quality negatively.

The authors of the study also noted a contradiction, their research shows that "time budget pressure" is influencing the quality of work for the worse, but not the well-being of the auditor, which calls for further studies. Similar results were also produced by surveying small audit firms [15]. On the other hand, unusually long audit times have been linked with a large number of errors in original reports [16], suggesting that fraudulent misstatements are detected less under time pressure [17]. Auditors specialized in a specific industry, experience less time pressure than non-specialized ones, however the size of the company being audited has no effect [18]. Experienced auditors make less errors under time pressure [19]. Time pressure had also been linked to under reporting of overtime [20].

In an experiment done by Bowrin and King [21], the U-shaped performance predicted by Yerkes Dodson's law could not be replicated in auditing tasks. Auditors performance did not decrease with low time pressure. The authors note that the tasks and its context in the research setting might not bring arousal down to a low level associated with decreased performance. Otherwise the authors detect that performance decreases with increased time pressure. Additionally, the performance decrease is greater in the more complex task of the two used in the study.

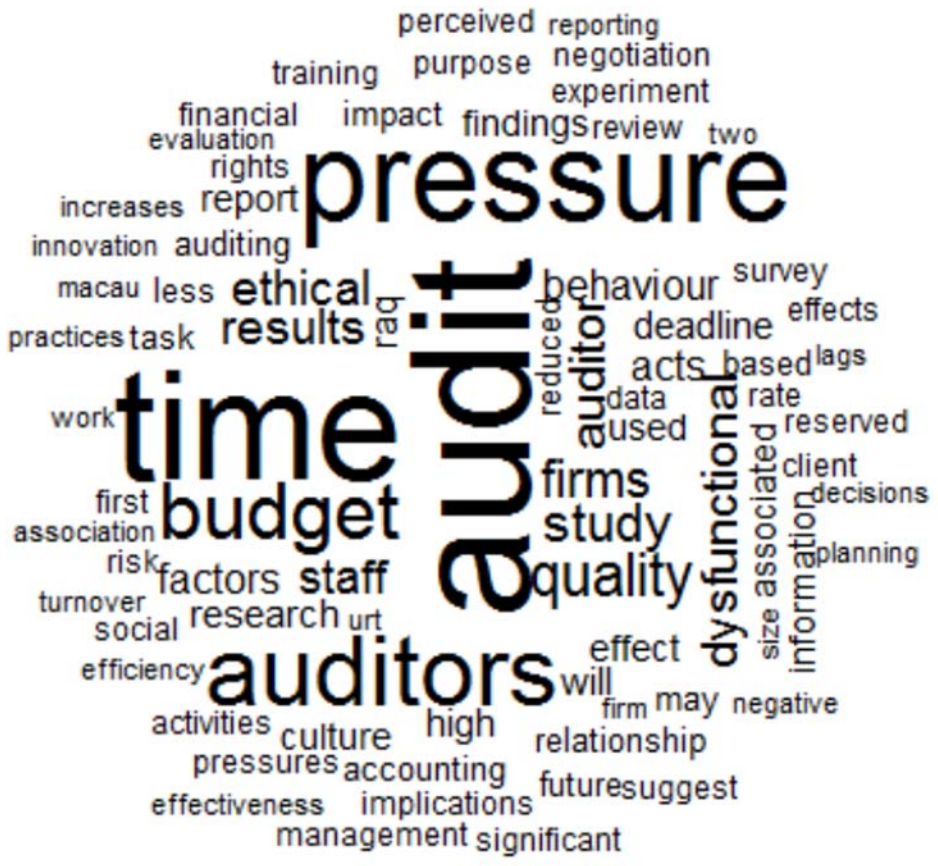

Figure 5. Wordcloud of topic "auditors" made from titles and abstracts.

There has also been research on the effects of time pressure on negotiations between auditors and their clients chief financial officers [22]. Based on surveys, it was deemed that auditors with time pressure conceded less in pre-negotiations, but were willing to concede more than their clients during negotiations.

Time pressure has also been linked to the ethicality of decisions in auditing [23]. This conclusion was based on questionnaires distributed in Ireland and USA. The country of the respondent had an effect on the results as the respondents from the USA reported overall more ethical practices. Results also indicate that ethical behavior was less affected by pressure created by management than by social pressure induced by peers. Hence the authors note that: "While management can expend considerable resources to establish and communicate the appropriate tone at the top, they must also find ways to ensure that regular work pressure experienced by staff does not conflict with the espoused values of the firm".

### D. *Job Satisfaction*

The topic with the five most probable words "job", "satisfaction", "physicians", "workers" and "organizational" was put under two high level nodes in qualitative coding: "Occupations" and "Wellbeing". Most of the 17 papers on this topic are related to Job Demands-Resources model [3].

In general, time pressure associates with job satisfaction negatively [24]. Surveys tell that child welfare supervisors who report time pressures also report more stress, while a direct link to intent leaving one's job could not be found [25].

Silla and Gamero [26] studied time pressure in the organizational level and call it "shared time pressure". In their study, 367 workers from 34 organizations in the road transport sector answered surveys on time pressure. Results indicate that time pressure has a negative effect on job satisfaction and self-reported health, but no connection to sickness and absence from workplace could be made. Possible explanation to this is the "shared time pressure": absence from work leads to increased time pressure for coworkers in the context of professional drivers, and thus the fear of retaliation might significantly discourage absence.

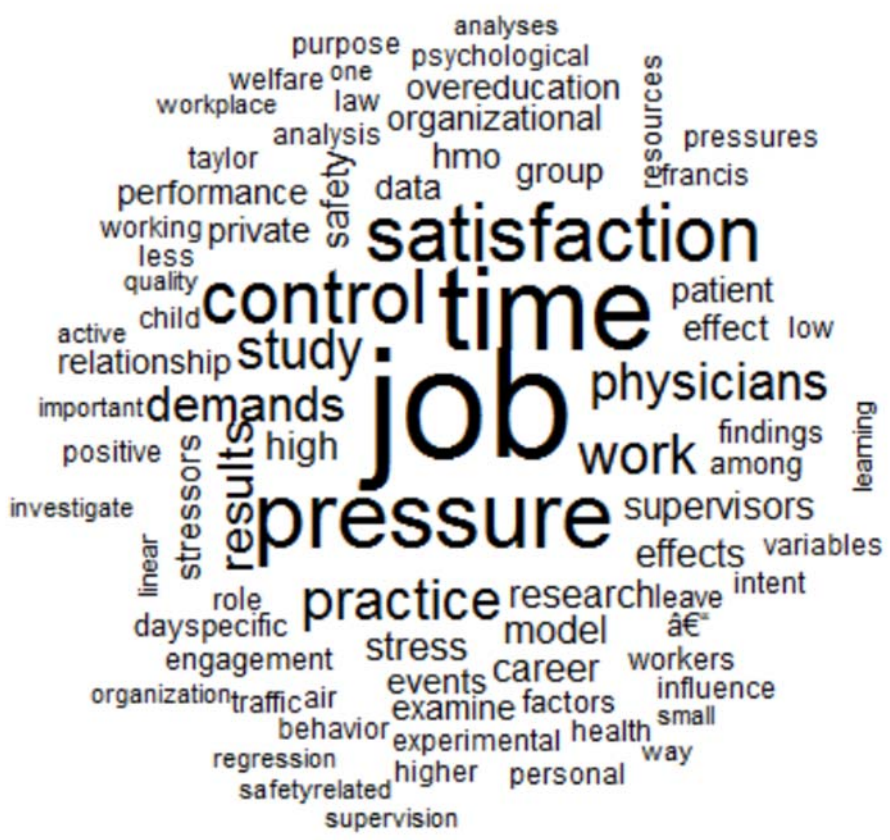

Figure 6. Wordcloud of "Job Satisfaction" made from titles and abstracts.

Häusser, Schultz-Hardt and Mojzisch [27] did an experiment by simulating office work and examining the speed-accuracy tradeoff. Examined variables were job demands, which was the number of tasks in a specific timeframe, and control, which meant there was pacing of the work. The authors used post-test, where measuring was done when the variables of demands and control were not manipulated, instead they were manipulated before the post-test to see effects on learning. Speed-accuracy tradeoff could be found post-test on control, workers with less control and more pacing worked slower, but their work was more accurate. However, speed-accuracy tradeoff could not be found post-test on high demand conditions, workers with high demands worked faster post-test without being less accurate. The authors offer learning effect as an explanation for these results.

Diary study by Kühnel, Sonnentag and Bledow [28] looked at day-level time pressure and other variables as antecedents to level of work engagement. They found out that time pressure was positively associated with work engagement. On days workers experienced less control over their work, they also reported less engagement. Hence authors deduce that it might be productive to grant more control to workers when they are under time pressure. Also, the authors report that work engagement varied significantly over time.

Time pressure has been investigated in engineering project work, where perceived time pressure had slight negative effects on job satisfaction and goal fulfillment [29]. The support and collective ability of the team were a negating factor for these effects.

Study examining the job satisfaction of physicians found out that those working for health maintenance organizations experienced more time pressure than other physicians (e.g. those working solo), while also having lower job satisfaction and reporting higher intentions to leave their practices [30].

## V. Discussion and Future Work

In this paper, we analyzed 1270 papers related to time pressure found on Scopus using clustering and qualitative coding. As a result of the literature review performed on the four topics obtained from clustering, a list of testable hypothesis is offered. These hypotheses might benefit from additional evidence or the evidence has been gathered in a different context than software engineering. Table 1 lists the hypothesis and the interdisciplinary sources from which they have been extracted. Conducting studies in order to assess their validity in the context of software development would give more insights on how time pressure affects software developers' productivity, emotions and mental health.

We are aware of some findings not included in the clusters of papers analyzed here and their relevance to the proposed hypotheses. Khomh, Dhaliwal, Zou and Adams [31] found out that Mozilla's Firefox project changing to rapid releases did not increase the number of bugs, but bugs described as "showstoppers" increased. We suspect that the workflow of the company and the nature of deadlines have an impact on the effects of time pressure, i.e. "softer" deadlines where new features can be delayed to the next release lower or erase the effects of time pressure.

**Table 1**

| **Hypothesis** | **Potential Operationalizations Sw** | **Origin** |
|---|---|---|
| Tight development processes support time management | | Hazzan and Dubinsky [9] |
| Communication lessens with pressure for project completion or deadline. | Less comments in bug tracker near deadlines. | Van Berkel, Ferguson and Groenewegen [10] |
| Tendency to overly focus on the technical solution | | Hussein[11] |
| Tendency to make unverified assumptions | Less test runs or less tester feedback | Hussein[11] |
| Significant rise to personal emotions, such as fear, diffidence, competitiveness and eagerness | Sentiment analysis of bug tracker messages. | Hussein [11] |
| Time Pressure increases errors and lessens quality. | More bugs are introduced to the code near deadlines. | Cataldo [12] |
| Time pressure increases errors in more complex tasks compared to simpler tasks. | More bugs introduced to more complex classes and functions near deadlines. | Bowrin and King II [21] |
| Projects overdue are lower in quality and need more rework. | Deadlines are postponed and/or more post-release bugs | Blankley, Hurtt and MacGregor [16] |
| Experienced workers are less affected by time pressure and workload | | Cianci and Bierstaker [19] |
| Giving worker increased control under time pressure increases productivity | | Kühnel, Sonnentag and Bledow [28] |
| Team ability and support negate effects of time pressure | | Nordqvist, Hovmark and Zika-Viktorsson [29] |
| Higher turnover of workforce in companies with higher time pressure | | Linzer et al. [30] |

There are some possibly conflicting reports in the overviewed literature, Bowrin [21] notices a bigger effect of time pressure on more complex tasks in auditing, while Häusser [27] does not find less accuracy with more demands in office work simulation. This is further pointing out towards the context dependent nature of time pressure.

As future work, we plan to mine different software repositories in order to test several of those hypotheses using metrics such as valence and arousal. Comparing the results obtained on different software projects would allow us to gain knowledge on what type(s) of time pressure policies positively affect both software developers' emotions and health, and thus software development. Moreover, this work can be further expanded by overviewing more topics found by clustering and adding bibliometric data. More papers about software engineering time pressure can be covered by performing snowballing from the identified sources.

## Acknowledments

The first, second and third author have been supported by Academy of Finland grant 298020.